\documentclass[a4paper]{article}

\usepackage{INTERSPEECH2021}
\usepackage{hyperref}
\usepackage{multirow}
\usepackage{flushend}

\title{The IDLAB VoxCeleb Speaker Recognition Challenge 2021 System Description}
\name{Jenthe Thienpondt\thanks{$^*$Equal contribution}$^*$, Brecht Desplanques$^*$, Kris Demuynck}

\address{
    IDLab, Department of Electronics and Information Systems, Ghent University - imec, Belgium
}
\email{jenthe.thienpondt@ugent.be, brecht.desplanques@ugent.be}

\begin{document}

\maketitle
\begin{abstract}
This technical report describes the IDLab submission for track~1 and 2 of the VoxCeleb Speaker Recognition Challenge 2021~(VoxSRC-21). This speaker verification competition focuses on short duration test recordings and cross-lingual trials.
Currently, both Time Delay Neural Networks (TDNNs) and ResNets achieve state-of-the-art results in speaker verification. We opt to use a system fusion of hybrid architectures in our final submission.
An ECAPA-TDNN baseline is enhanced with a 2D convolutional stem to transfer some of the strong characteristics of a ResNet based model to this hybrid CNN-TDNN architecture. Similarly, we incorporate absolute frequency positional information in the SE-ResNet architectures.
All models are trained with a special mini-batch data sampling technique which constructs mini-batches with data that is the most challenging for the system on the level of intra-speaker variability. This intra-speaker variability is mainly caused by differences in language and background conditions between the speaker's utterances. The cross-lingual effects on the speaker verification scores are further compensated by introducing a binary cross-linguality trial feature in the logistic regression based system calibration. The final system fusion with two ECAPA CNN-TDNNs and three SE-ResNets enhanced with frequency positional information achieved a third place on the VoxSRC-21 leaderboard for both track 1 and 2 with a minDCF of 0.1291 and 0.1313 respectively.


\end{abstract}

\section{System architectures}

TDNN and ResNet based speaker verification systems achieve similar performance~\cite{voxsrc_2020_technical_report} while being structurally very different. TDNNs depend on 1D convolutions of which the kernels cover the complete frequency range of the input features. As a consequence, the absolute frequency position of each input feature is hard-coded through the order of the filter coefficients. This makes sense in the context of speaker verification as features based on absolute frequency information such as the speaker's pitch provide crucial information. The main drawback is that many filters are needed to model the fine details of complex patterns that can occur at any frequency region. 


ResNets use 2D convolutions with a small receptive field, capturing local frequency and temporal information. By exploiting local speaker-specific frequency patterns that repeat but for small frequency shifts, this model requires fewer feature channels to model fine frequency details. However, accurate absolute frequency positions of features are not explicitly encoded~\cite{position_info_resnet}. This can be sub-optimal as we expect significantly different patterns in the low frequency regions compared to the high frequency regions. 
We enhance both architectures to incorporate the beneficial characteristics of the other model. The modifications are described below and more information can be found in~\cite{ecapa_cnn_tdnn}.

\begin{figure}[ht]
\centering
\includegraphics[scale=0.32]{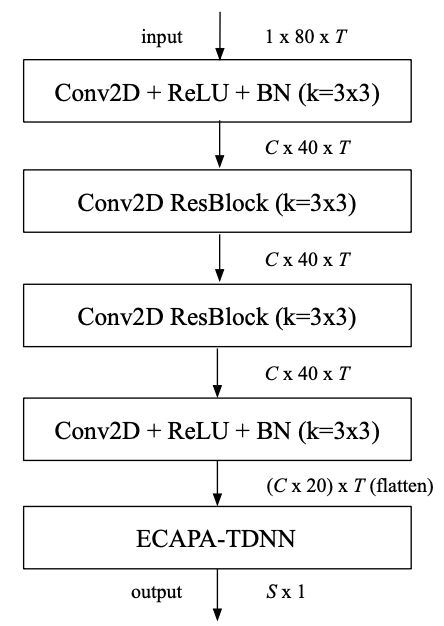}
\caption{{\it The 2D convolutional stem of the ECAPA CNN-TDNN architecture. $C$ and $T$ indicate the channel and temporal dimension, respectively. The kernel size is denoted by $k$. $S$ refers to the number of training speakers.}}
\label{fig:conv_stem}
\end{figure}

\subsection{ECAPA CNN-TDNN with 2D convolutional stem}

Architectures based on the TDNN topology apply the 1D-convolutional operation on the frame-level intermediate features. Consequentially, the kernel of the convolution encompasses the complete frequency range.
We want the ECAPA-TDNN~\cite{ecapa_tdnn} to be equivariant to small and reasonable shifts in the frequency domain, compensating for realistic intra-speaker frequency variability. To accomplish this, we base the initial network layers on 2D convolutions. These layers will also require less filters to model high resolution input detail.

Inspired by the powerful 2D convolutions in the ResNet architectures, we adopt a similar structure for our ECAPA-TDNN stem. The proposed network configuration is shown in Figure~\ref{fig:conv_stem}. 
To make the stem more computationally efficient we use a stride of 2 in the frequency dimension of the first and final 2D convolutional layer. The output feature maps of the new stem are subsequently flattened in the channel and frequency dimensions and used as input for the regular ECAPA-TDNN network. We will refer to this network as the ECAPA CNN-TDNN.

\subsection{SE-ResNet with frequency position information}
\label{section:fwse_resnet}

The second architecture is based on the ResNet~\cite{resnet} architecture. We use the same network topology as defined in~\cite{magneto}, with the addition of an SE-block~\cite{se_block} at the end of each residual block. We also incorporate the same channel-dependent attentive statistics pooling layer as in~\cite{ecapa_tdnn}.

\subsubsection{Learnable frequency positional encodings}

While a certain robustness against frequency translation is beneficial, the spatial equivariance of 2D convolutions can limit the ability of the network to fully exploit frequency-specific information. We argue that encoding positional frequency information in the intermediate feature representations can make the network incorporate and utilize knowledge of the feature frequency positions.

Positional encodings have been popularized with the rise of Transformer~\cite{transformer} models. By design, these Transformers are invariant to the re-ordering of the input sequence and positional encodings can alleviate this issue when required. 
In our submission we focus on learnable encodings as the computational overhead is negligible.

Consider the input of a residual module in our baseline ResNet architecture as $ \pmb{X} \in \mathbb{R}^{C \times F \times T} $,  with $C$, $F$ and $T$ indicating the channel, frequency and time dimension, respectively.
We define the positional encoding vector $\pmb{p} \in \mathbb{R}^{F} $ as a trainable vector.
The elements in this vector are broadcasted to match the dimensions of the targeted input feature maps. The input of the residual module is now defined as $\pmb{X} + \pmb{p}$. We add a unique positional encoding to the input of each residual block after branching the skip connection. 


\subsubsection{Frequency-wise Squeeze-Excitation (fwSE)}

Squeeze-Excitation (SE)~\cite{se_block} blocks have been successfully applied in both TDNN and ResNet based speaker verification architectures. The first stage, the squeeze operation, calculates a vector $ \pmb{z} \in \mathbb{R}^{C} $ containing the mean descriptor for each feature map. It is followed by the excitation operation, which calculates a scalar for each feature map given the information in $\pmb{z}$. Subsequently, each feature map is rescaled with its corresponding scalar.

We argue that rescaling per feature map is not tailored for the processing of speech in ResNets. Instead, we propose a frequency-wise Squeeze-Excitation (fwSE) block, which injects global frequency information across all feature maps. We calculate a vector $ \pmb{z} \in \mathbb{R}^{F} $ containing the mean descriptor for each frequency-channel of the intermediate feature maps in the following manner:
\begin{equation}
\label{channel_se}
z^{f} = \frac{1}{C \times T} \sum_{i=1}^{C} \sum_{j=1}^{T} x^{f}_{i,j}
\end{equation}
with $x^f$ the elements of $\pmb{X}^{f} \in \mathbb{R}^{C \times T}$, the component of input feature map $ \pmb{X} \in \mathbb{R}^{C \times F \times T} $ corresponding with frequency position $f$.
From the mean descriptors in $ \pmb{z} $ we calculate a vector $ \pmb{s} \in \mathbb{R}^F $ containing the scaling scalars for each frequency-channel in the second stage, the excitation operation:
\begin{equation}
\label{se_excitate}
\pmb{s} = \sigma( \pmb{W}_2 f(\pmb{W}_1 \pmb{z} + \pmb{b}_{1}) + \pmb{b}_2)
\end{equation}
with $ \pmb{W} $ and $ \pmb{b} $ indicating the weights and bias of a linear layer, $ f(.) $ the ReLU activation function and $ \sigma $ the sigmoid function. Finally, $\pmb{X}^f$ is scaled with the corresponding scalar in $\pmb{s}$. The proposed frequency-wise SE-blocks are inserted at the end of each residual module before the additive skip connection.

\subsection{Single system configurations}

The two selected ECAPA CNN-TDNN architectures in the system fusion use the same model architecture as described in~\cite{ecapa_cnn_tdnn}. We also introduce a larger variant \textit{ECAPA CNN-TDNN (big)} with 256 channels in the 2D convolutional stem compared to 128 channels in the original model. The number of channels in the TDNN layers is set to 2048 and the ECAPA-TDNN includes a total of four SE-Res2Blocks.

In addition, we train three fwSE-ResNet variants with a topology as described in Section~\ref{section:fwse_resnet}. We vary the amount of layers in each of the four ResBlocks of the model as indicated by the four numbers between brackets in Table~\ref{tab:single_system_results}.

\section{Training procedure}

\subsection{Initial training}

All speaker embedding extractors are trained on the development part of the VoxCeleb2 corpus~\cite{vox2}. The input features are 80-dimensional log Mel-filterbank energies extracted with a window length of 25 ms and a frame-shift of 10 ms. We apply an online augmentation strategy on random training crops using the MUSAN \cite{musan} corpus (additive music, noise \& babble) and the RIR \cite{rirs} corpus (reverb). SpecAugment \cite{specaugment} is applied to the input features with a frequency and temporal masking dimension of 10 and 5, respectively. Finally, the input features are mean normalized across time per utterance.

During the construction of each mini-batch we randomly select 32 utterances from different speakers and apply the four augmentations on random crops with a duration of 2\,s to get a total of 160 training samples. We also include a crop of the original clean signal in the mini-batch. SpecAugment is applied on all samples. This mini-batch data sampling procedure has been shown to be beneficial for speaker diarization performance in~\cite{ecapa_diarization}.

The model parameter updates are determined by the Adam optimizer \cite{adam} with a cyclical learning rate schedule. We use the \textit{triangular2} policy described in \cite{clr}. The cycle length is set to 130k iterations with a minimum and maximum learning rate of 1e-8 and 1e-3, respectively. The ECAPA CNN-TDNN systems are trained for three full cycles, while the fwSE-ResNets are trained for one cycle to prevent overfitting. The AAM-softmax~\cite{arcface} layers uses a margin $m$ of 0.2 and a scale $s$ of 30. To prevent overfitting, a weight decay of 2e-4 is applied on the weights of the AAM-softmax layer, while a value of 2e-5 is used on all other layers of the model.



\subsection{Large margin fine-tuning and cross-lingual sampling}
\label{sec:lm_ft}

We fine-tune all systems with an adapted version of the Large Margin Fine-Tuning (LM-FT) strategy presented in \cite{icassp_voxsrc20}. The CNN-TDNN systems use the original settings with an increase in crop size to 6 seconds accompanied with a margin increase to 0.5. For computational reasons, we limited the increase in crop length to only 4 seconds with a margin of 0.4 for the ResNets systems during fine-tuning.

To make the embeddings of the models more robust against variations in language of the same speaker, we used an alternative training utterance sample technique to construct challenging cross-lingual mini-batches during fine-tuning. Each mini-batch contains two utterances of a different language for 64 training speakers resulting in a total mini-batch size of 128. This results in more robustness of the embeddings against the challenging cross-lingual test trial conditions. To determine the language of the utterances we use the estimated language labels provided by the VoxSRC-21 organizers. Each utterance is optionally augmented with music, babble, noise, reverb or clean with equal probability.

Due to time constraints, we did not investigate the possibility to combine the hard prototype mining based utterance selection criterion~\cite{sdsvc_paper, icassp_voxsrc20} with the cross-lingual sampling.

\section{Scoring procedure}

\subsection{Score estimation and normalization}
\label{sec:score_norm}

The verification trials are scored by calculating the cosine similarity between the enrollment model and the test utterance speaker embeddings. The scores are normalized using top-100 adaptive s-normalization \cite{as_norm_original_1, as_norm_original_2}. The imposter cohort consists of top imposters selected from the speakers in the VoxCeleb2 training set. Each imposter speaker is represented by the average of their length-normalized training embeddings.

\subsection{Score fusion and QMF score calibration}
\label{section:QMF}

We fuse the systems scores by taking the weighted average of the s-normalized single system scores. In line with the single system performance on the validation set, the ResNet based systems get assigned double the weight of the TDNN systems. Next, we perform quality-aware score calibration~\cite{icassp_voxsrc20} based on logistic regression with Quality Metric Functions (QMFs). We use the minimum and maximum value of $\log(d_u - d_{min})$ for the enrollment and test utterance as a QMF with $d_u$ the duration of the considered utterance and $d_{min}$ slightly lower then the minimum expected test utterance duration (i.e.\ 1.99 s). A second set of QMFs is the minimum and maximum imposter mean score of both utterances in the trial against an imposter cohort consisting of the average utterance embedding of the VoxCeleb2 training speakers. As analyzed in~\cite{voxsrc_2020_technical_report}, we use the inner product instead of the cosine similarity to compute this score as this results in better performance. Finally, we add a binary cross-linguality feature in the calibration backend indicating if the enrollment and test utterance are spoken in the same language or not according to the provided language label.

For the open condition, we use the provided validation set to train the logistic regression calibration backend. To make the calibration system more robust against short duration trial conditions, we randomly crop half of the utterances included in the validation set between 2 and 4 seconds.

For the closed track, we create a validation set based on utterances of the VoxCeleb2 development set. We create 100K within-gender and cross-video trials, with an even split of positive and negative trials. Again, half of the utterances are cropped between 2 and 4 seconds to simulate short duration conditions. Half of the trials are cross-lingual to make the system calibration robust against varying language conditions. Because this calibration set uses the same utterances as those on which the models are trained, the impact of cross-linguality conditions will be underestimated by the calibration backend. To compensate, we consider the weight of the binary cross-linguality feature as a hyper-parameter to tune manually according to the performance on the VoxSRC-21 validation set.

\section{Results}



We evaluate all single systems and (calibrated) fusion system on the given VoxSRC-21 validation set in Table~\ref{tab:single_system_results}. The results of the calibrated system fusion on the VoxSRC-21 test set is provided in Table~\ref{tab:fusion_results}. We report the EER and MinDCF metric using a $P_{target}$ value of $0.05$ with $C_{Miss} = 1$ and $C_{FA} = 1$. All reported scores are s-normalized using an imposter cohort size of 100 speakers. The single system scores and system fusion without QMFs in Table~\ref{tab:single_system_results} are not calibrated, while the reported fusion scores with QMFs are calibrated as described in Section~\ref{section:QMF}.

\begin{table}[h]
    \caption{System performance on the VoxSRC-21 validation set.}
    \label{tab:single_system_results}
  \centering
  \begin{tabular}{lcc}
    \toprule
        \multicolumn{1}{c}{\textbf{System}} &
        \multicolumn{1}{c}{\textbf{EER(\%)}} &
        \multicolumn{1}{c}{\textbf{MinDCF}} \\
    \midrule
    ECAPA CNN-TDNN & 2.67 & 0.1604 \\
    ECAPA CNN-TDNN (big) & 2.65 & 0.1531 \\
    fwSE-ResNet86 (24, 32, 24, 6) & 2.37 & 0.1296 \\ 
    fwSE-ResNet110 (24, 48, 32, 6) & 2.48 & 0.1334 \\ 
    fwSE-ResNet90 (12, 32, 40, 6) & 2.45 & 0.1305 \\ 
    \midrule
    Fusion & 1.71 & 0.0986 \\ 
    Fusion + QMFs (closed) & 1.54 & 0.0943 \\ 
    Fusion + QMFs (open) & \textbf{1.38} & \textbf{0.0880} \\ 
    \bottomrule
  \end{tabular}
\end{table}

\begin{table}[h]
    \caption{Fusion system performance on the VoxSRC-21 test set.}
    \label{tab:fusion_results}
  \centering
  \begin{tabular}{lcc}
    \toprule
        \multicolumn{1}{c}{\textbf{Systems}} &
        \multicolumn{1}{c}{\textbf{EER(\%)}} &
        \multicolumn{1}{c}{\textbf{MinDCF}} \\
    \midrule
    Fusion + QMFs (closed) & 2.27 & \textbf{0.1291} \\
    Fusion + QMFs (open) & \textbf{2.05} & 0.1313 \\ 
    \bottomrule
  \end{tabular}
\end{table}

\bibliographystyle{IEEEtran}

\bibliography{mybib}


\end{document}